\documentclass[useAMS,usenatbib]{mn2e}

\usepackage{graphicx}
\usepackage{amsmath}
\usepackage{amssymb}
\usepackage{color}
\usepackage{mathtools}

\usepackage[breaklinks,colorlinks,citecolor=blue,linkcolor=red]{hyperref} 
\usepackage[all]{hypcap}

\newcommand\msun{\, \rm M_\odot}

\newcommand\kms{\, \rm km\,s^{-1}}

\newcommand\be{\begin{equation}}
\newcommand\ee{\end{equation}}

\title[Recoil kicks for BH mergers in LIGO/Virgo catalogs]{Implications of recoil kicks for black hole mergers from LIGO/Virgo catalogs}
\author[G. Fragione \& A. Loeb]{\parbox{\textwidth}{Giacomo Fragione$^{1,2}$\thanks{E-mail: giacomo.fragione@northwestern.edu}, Abraham Loeb$^{3}$}\\
\ \\
$^1$Department of Physics \& Astronomy, Northwestern University, Evanston, IL 60202, USA\\
$^2$Center for Interdisciplinary Exploration \& Research in Astrophysics (CIERA), Evanston, IL 60202, USA\\
$^3$Astronomy Department, Harvard University, 60 Garden St., Cambridge, MA 02138, USA}

\begin{document}

\maketitle

\begin{abstract}
The first and second Gravitational Wave Transient Catalogs by the LIGO/Virgo Collaboration include $50$ confirmed merger events from the first, second, and first half of the third observational runs. We compute the distribution of recoil kicks imparted to the merger remnants and estimate their retention probability within various astrophysical environments as a function of the maximum progenitor spin ($\chi_{\rm max}$), assuming that the LIGO/Virgo binary black hole (BBH) mergers were catalyzed by dynamical assembly in a dense star cluster. We find that the distributions of average recoil kicks are peaked at about $150\kms$, $250\kms$, $350\kms$, $600\kms$, for maximum progenitor spins of $0.1$, $0.3$, $0.5$, $0.8$, respectively. Only environments with escape speed $\gtrsim 100\kms$, as found in galactic nuclear star clusters as well as in the most massive globular clusters and super star clusters, could efficiently retain the merger remnants of the LIGO/Virgo BBH population even for low progenitor spins ($\chi_{\rm max}=0.1$). In the case of high progenitor spins ($\chi_{\rm max}\gtrsim 0.5$), only the most massive nuclear star clusters can retain the merger products. We also show that the estimated values of the effective spin and of the remnant spin of GW170729, GW190412, GW190519\_153544, and GW190620\_030421 can be reproduced if their progenitors were moderately spinning ($\chi_{\rm max}\gtrsim 0.3$), while for GW190517\_055101 if the progenitors were rapidly spinning ($\chi_{\rm max}\gtrsim 0.8$). Alternatively, some of these events could be explained if at least one of the progenitors is already a second-generation BH, originated from a previous merger.
\end{abstract}

\begin{keywords}
stars: black holes -- galaxies: kinematics and dynamics -- stars: black holes -- stars: kinematics and dynamics -- galaxies: nuclei
\end{keywords}

\section{Introduction}
\label{sect:intro}

The LIGO/Virgo Collaboration has recently released the second Gravitational Wave Transient Catalog (GWTC-2), which includes events from the first half of the third observational run. Together, GWTC-1, from the first two observational runs \citep{lvc2019cat}, and the new candidate events presented in \citep{lvc2020cat} comprise GWTC-2, which includes $50$ events, that are revolutionizing our understanding of black holes (BHs) and neutron stars (NSs). Thanks to the growing number of detected events, the distributions of masses, spins, and merger rates can be constrained with unprecedented statistical precision and GW events provide a unique opportunity to probe fundamental physics \citep{lvc2020catb,lvc2020catc}.

The origin of binary mergers is still highly uncertain, with several possible scenarios that could potentially account for most of the observed events  \citep[e.g.,][]{antoper12,belcz2016,askar17,bart17,giac2018,ll18,baner18,fragk2018,rod18,fragg2019,fragk2019,ham2019,krem2019,rasskoc2019,fragrasio2020}. While several models account for roughly the same rate, the statistical contribution of different astrophysical channels can be hopefully disentangled using a combination of the mass, spin, redshift, and eccentricity distributions \citep[e.g.,][]{olea09,Fishbachetal2017,gondan2018,perna2019}.

Amongst the GWTC-2 systems, there are high-mass BBHs, the most massive of which being the source of GW190521 \citep{ligo2020new1}. BHs in the mass range of about $45$--$135\msun$ are not expected to form via standard stellar evolution since the pair-instability process either limits the maximum mass of the core of the progenitor star or disrupts it entirely \citep[e.g.,][]{heger2003,woosley2017}.

A natural way to form BHs in (and above) the pair-instability mass gap is through hierarchical mergers in a dense stellar environment, where the remnant of a previous merger becomes part of a new binary \citep[e.g.,][]{anto2019,frsilk2020,mapell2020}. Other processes include stellar mergers, formation of BHs from Population III stars, or growth via accretion in an AGN disk \citep[e.g.,][]{kinug2020,krem2020,tagawa2020,vanson2020}. The main barrier to the formation of second- (or higher-) generation BHs via hierarchical mergers stems from the recoil kick imparted to the merger remnant as a result of anisotropic GW emission \citep{lou10,lou12}, which could eject it from the parent stellar cluster \citep*[e.g.,][]{fraga2018,fragb2018}. Therefore, the relative magnitude of the cluster escape speed, set by its mass and density, compared to the magnitude of the recoil kick determines the maximum mass achievable through repeated mergers.

Assuming that LIGO/Virgo BBH mergers were catalyzed by dynamical assembly and interactions in a dense star cluster \citep[e.g.,][]{wong2020}, we compute the distribution of recoil kicks imparted to their merger remnants and estimate their retention probability within various astrophysical environments. If the retention probability is high enough, the remnants of LIGO/Virgo BBH mergers could eventually form a new binary and merge again \citep[e.g.,][]{anto2019,frsilk2020,mapell2020}.

The paper is organized as follows. In Section~\ref{sect:method}, we describe the method we use to compute the recoil kicks imparted to merger remnants from anisotropic gravitational wave emission. In Section~\ref{sect:results}, we present the results of our calculations. Finally, in Section~\ref{sect:concl}, we draw our conclusions.

\section{Method}
\label{sect:method}

\begin{figure} 
\centering
\includegraphics[scale=0.55]{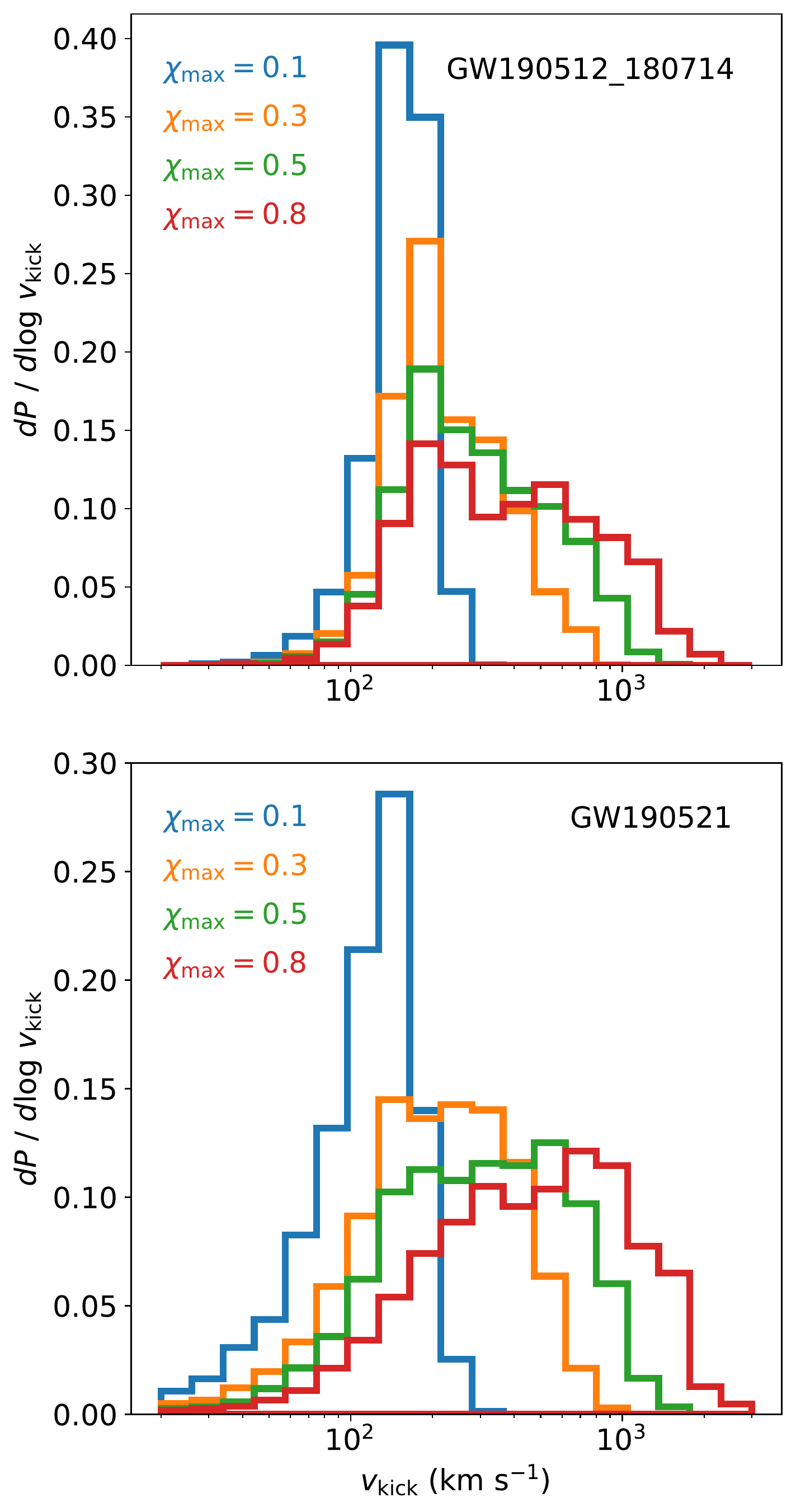}
\caption{Probability distribution function of the recoil kick ($v_{\rm kick}$) imparted to the merger remnant of GW190512\_180714 (top) and GW190521 (bottom). Different colors represent different values of $\chi_{\rm max}$: $0.1$ (blue), $0.3$ (orange), $0.5$ (green), $0.8$ (red).}
\label{fig:example}
\end{figure}

\begin{figure} 
\centering
\includegraphics[scale=0.55]{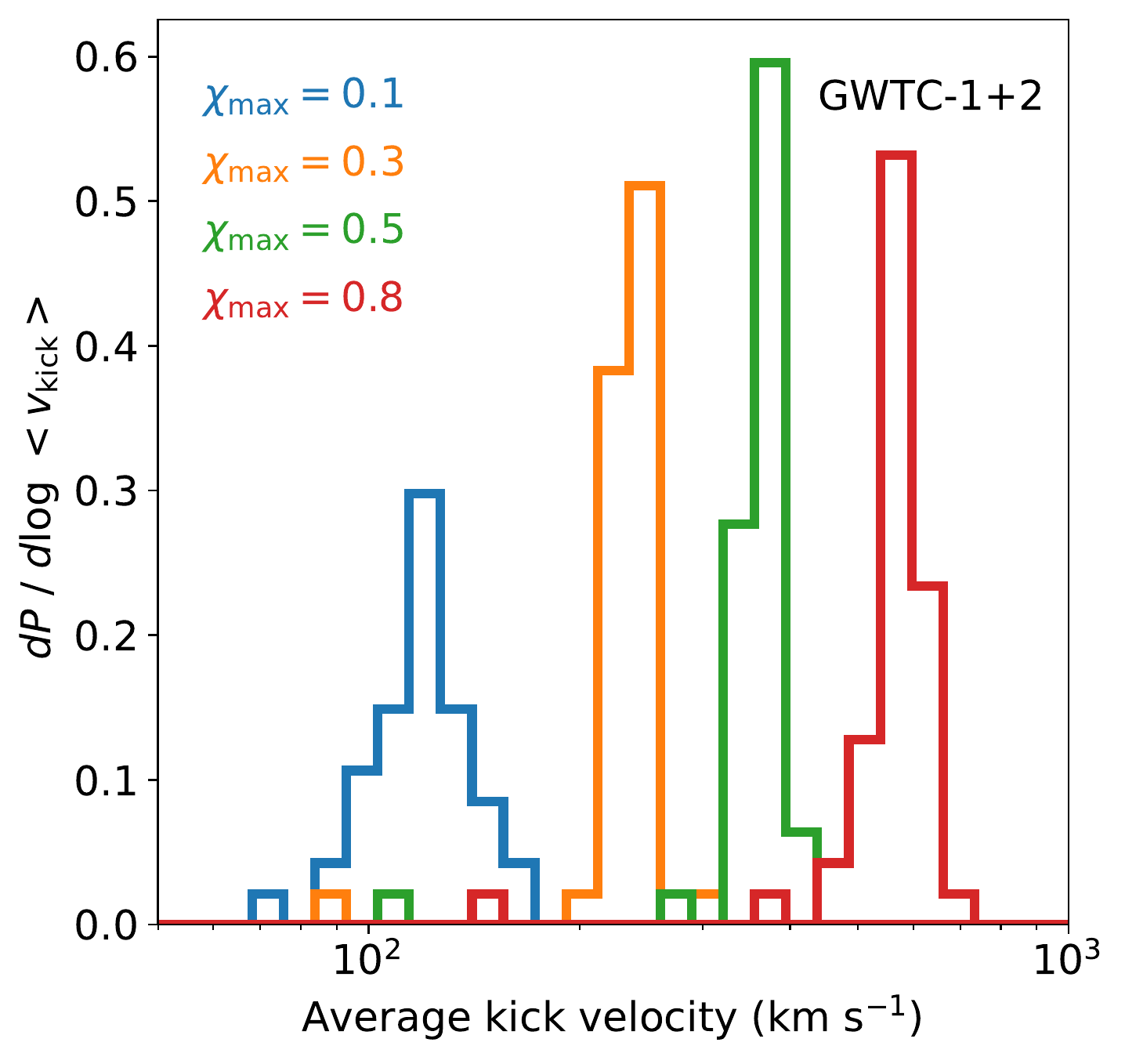}
\caption{Distribution of average kick velocity for the LIGO/Virgo population in GWTC-1 \citep{lvc2019cat} and GWTC-2 \citep{lvc2020cat} as a function of different values of $\chi_{\rm max}$.}
\label{fig:all}
\end{figure}

We model the recoil kick following \citet{lou10,lou12}
\begin{equation}
\boldsymbol{v}_{\mathrm{kick}}=v_m \hat{e}_{\perp,1}+v_{\perp}(\cos \xi \hat{e}_{\perp,1}+\sin \xi \hat{e}_{\perp,2})+v_{\parallel} \hat{e}_{\parallel}\,,
\label{eqn:vkick}
\end{equation}
where
\begin{eqnarray}
v_m&=&A\eta^2\sqrt{1-4\eta}(1+B\eta)\\
v_{\perp}&=&\frac{H\eta^2}{1+q}(\chi_{2,\parallel}-q\chi_{1,\parallel})\\
v_{\parallel}&=&\frac{16\eta^2}{1+q}[V_{1,1}+V_A \tilde{S}_{\parallel}+V_B \tilde{S}^2_{\parallel}+V_C \tilde{S}_{\parallel}^3]\times \nonumber\\
&\times & |\mathbf{\chi}_{2,\perp}-q\mathbf{\chi}_{1,\perp}| \cos(\phi_{\Delta}-\phi_{1})\,.
\end{eqnarray}
In the previous equations, $\eta=q/(1+q)^2$ is the symmetric mass ratio, $q=m_2/m_1<1$ is the binary mass ratio, $m_1$ and $m_2$ are the masses of the merging BHs, and $|\boldsymbol{{\chi_1}}|$ and $|\boldsymbol{{\chi_2}}|$ the magnitudes of their dimensionless spins. The $\perp$ and $\parallel$ refer to the directions perpendicular and parallel to the orbital angular momentum, respectively. Finally, $\hat{e}_{\perp,1}$ and $\hat{e}_{\perp,2}$ are orthogonal unit vectors in the orbital plane. We also define
\begin{equation}
\tilde{\boldsymbol{S}}=2\frac{\boldsymbol{\chi}_{2}+q^2\boldsymbol{\chi}_{1}}{(1+q)^2}\,,
\end{equation}
$\phi_{1}$ as the phase angle of the binary at merger, which we take random, and $\phi_{\Delta}$ as the angle between the in-plane component of the vector
\begin{equation}
\boldsymbol{\Delta}=M^2\frac{\boldsymbol{\chi}_{2}-q\boldsymbol{\chi}_{1}}{1+q}\,,
\end{equation}
where $M$ is the total binary mass and the infall direction at merger of the two BHs, which we sample uniformly. We adopt $A=1.2\times 10^4$ km s$^{-1}$, $H=6.9\times 10^3$ km s$^{-1}$, $B=-0.93$, $\xi=145^{\circ}$ \citep{gon07,lou08}, and $V_{1,1}=3678$ km s$^{-1}$, $V_A=2481$ km s$^{-1}$, $V_B=1793$ km s$^{-1}$, $V_C=1507$ km s$^{-1}$ \citep{lou12}. We compute the final total spin of the merger product and its mass following \citet{rezzolla2008}. 

\begin{figure*} 
\centering
\includegraphics[scale=0.625]{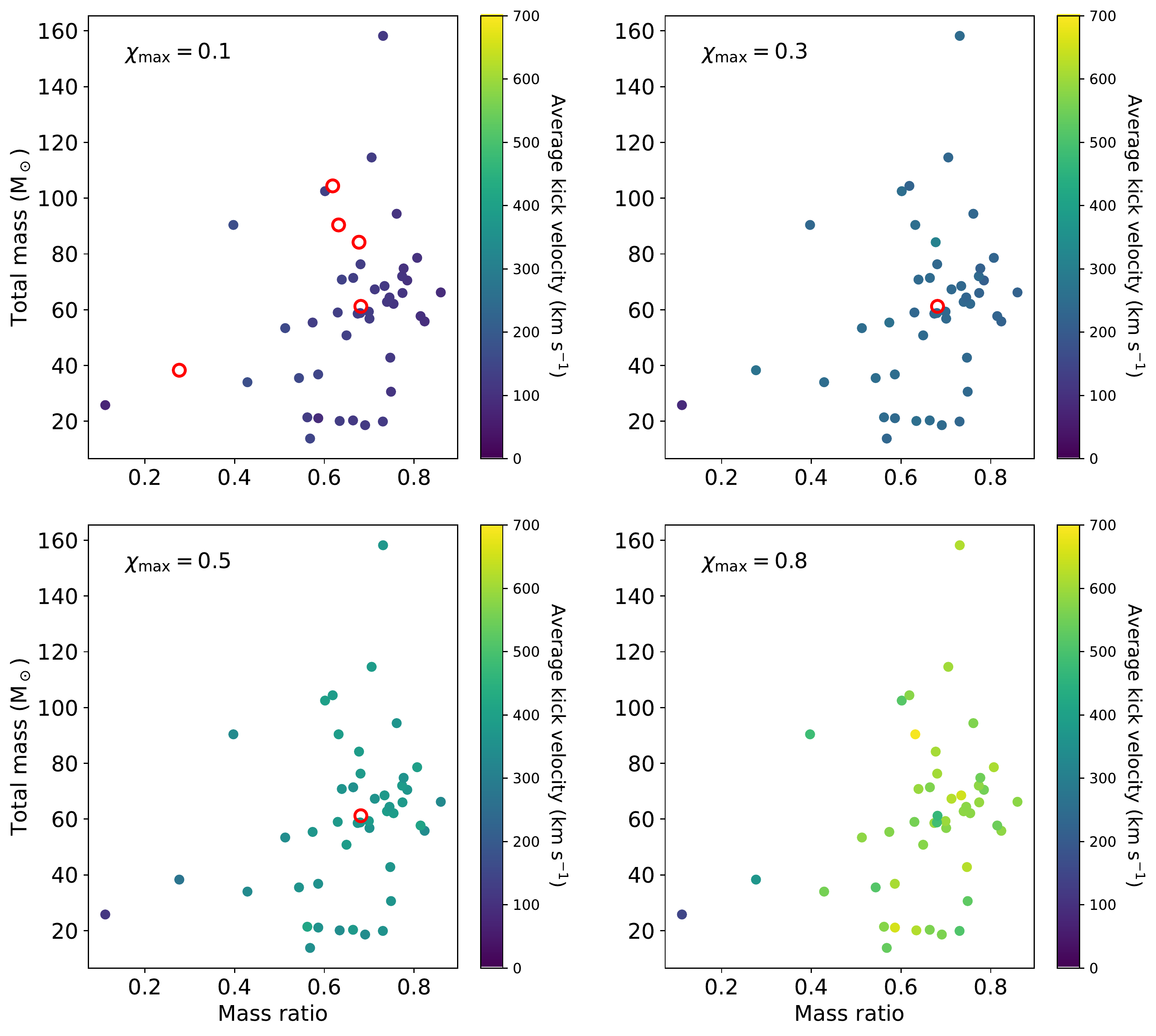}
\caption{Total mass as a function of the mass ratio for the BBH mergers in the LIGO/Virgo catalogs GWTC-1 \citep{lvc2019cat} and GWTC-2 \citep{lvc2020cat}. Color code: average kick velocity imparted to the merger remnant of BBH mergers from anisotropic gravitational wave emission. Different panels show different assumptions of $\chi_{\rm max}$: $0.1$ (top-left), $0.3$ (top-right), $0.5$ (bottom-left), $0.8$ (bottom-right). Red circles represent events whose $\chi_{\rm eff}$ and $\chi_{\rm fin}$ cannot reproduce the LIGO/Virgo estimated range for a given $\chi_{\rm max}$.}
\label{fig:mqvkick}
\end{figure*}

\begin{figure*} 
\centering
\includegraphics[scale=0.47]{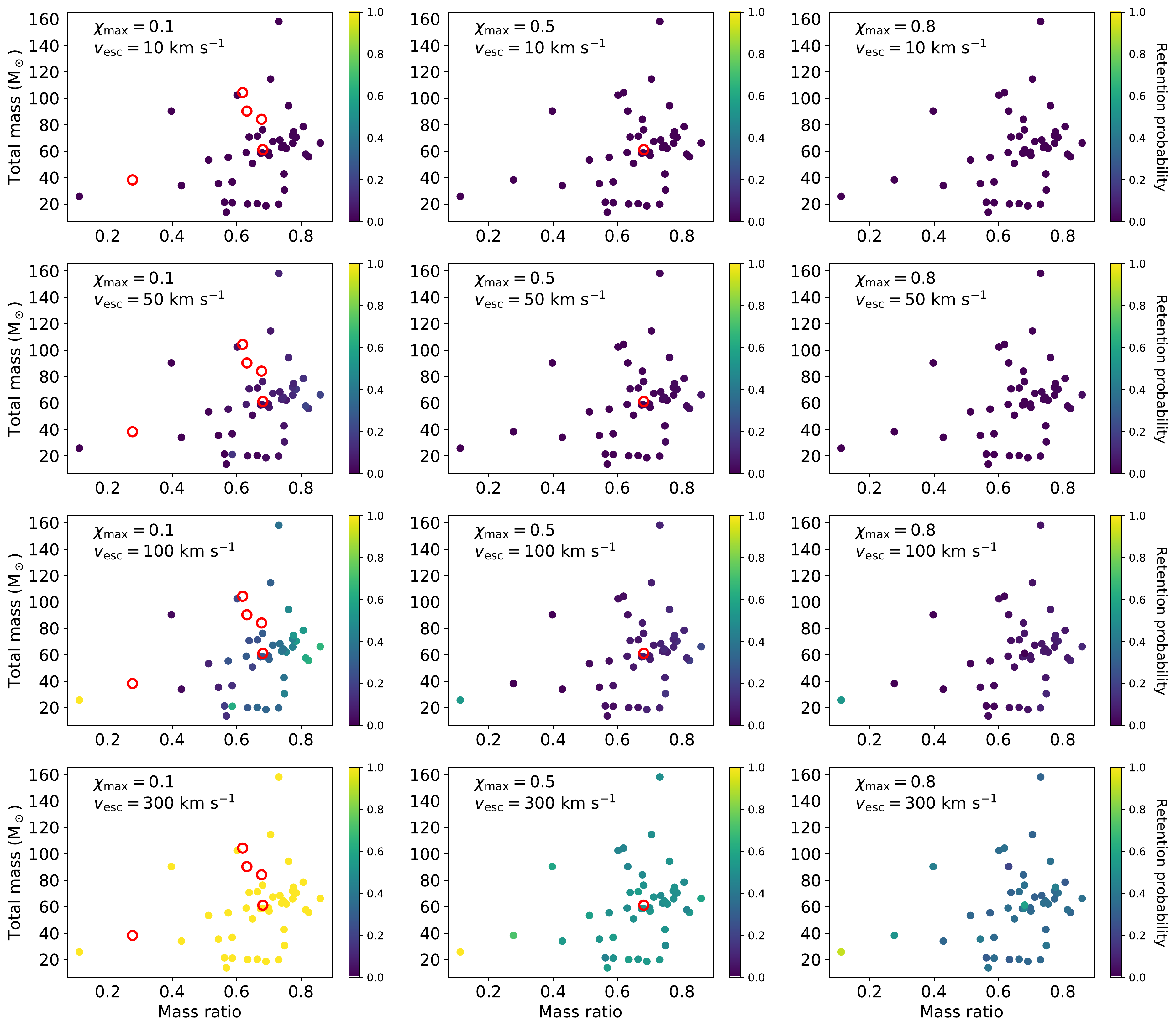}
\caption{Total mass as a function of the mass ratio for the BBH mergers in the LIGO/Virgo catalogs GWTC-1 \citep{lvc2019cat} and GWTC-2 \citep{lvc2020cat}. Color code: probability to retain the merger remnant of BBH mergers in environments with escape speed $v_{\rm esc}$. Different panels show different assumptions of $\chi_{\rm max}$: $0.1$ (left), $0.5$ (center), $0.8$ (right)}. Red circles represent events whose $\chi_{\rm eff}$ and $\chi_{\rm fin}$ cannot reproduce the LIGO/Virgo estimated range for a given $\chi_{\rm max}$.
\label{fig:mqprobret}
\end{figure*}

To compute the recoil kick imparted to the remnants of LIGO/Virgo BBH mergers, we consider the estimated masses $m_1$ and $m_2$ reported in GWTC-1 and GWTC-2. We sample the spins of the merging BHs uniformly in the range $[0,\chi_{\rm max}]$ and assume isotropic distribution of spins of each BH as proper to a dynamical environment. We run $10^5$ Monte Carlo simulations and compute the recoil kicks, using Eq.~\ref{eqn:vkick}. We discard all the simulations whose computed effective spin $\chi_{\rm eff}$ and spin of the merger remnant $\chi_{\rm fin}$ are not in the range allowed by LIGO/Virgo.

While rather simple, our method gives an estimate of the average magnitude of the recoil kick imparted to LIGO/Virgo merger remnants and their retention probability within various dynamical environments\footnote{We note that we are also neglecting possible correlations (if any) that could be present in the LIGO/Virgo multidimensional posteriors.}. Anyways, our kick computations remain astrophysically interesting since can be used as a proxy to understand the probability of formation of second-generation mergers in various environments, assuming the LIGO-Virgo events were catalyzed by dynamics. Other authors have proposed and developed more robust estimates of the recoil kicks based on the analysis of the gravitational waveform \citep{GerosaMoore2016,ligo2020new1,VarmaIsi2020}. However, they show that little information can be gained about the kick for existing events, and interesting measurements will soon become possible as detectors improve and space-based observatories become operative.

\section{Results}
\label{sect:results}

We use our framework to compute the recoil kick imparted to the merger remnants in the catalogs. In Figure~\ref{fig:example}, we show the probability distribution function (PDF) of the recoil kick ($v_{\rm kick}$) imparted to the merger remnant of GW190512\_180714 (top) and GW190521 (bottom), for different values of $\chi_{\rm max}$. GW190512\_180714 and GW190521 have mass ratio of about $0.54$ and $0.73$, respectively. The higher the spin, the larger the recoil kick imparted to the remnant and the broader its distribution. We find that the PDF for both events is peaked at $\sim 100\kms$ for $\chi_{\rm max}=0.1$, whereas it extends up to $\sim 1000\kms$ for $\chi_{\rm max}=0.8$. The merger remnant of GW190521 receives a larger kick as the progenitors' spins increase. Using waveform modeling and applying the related remnant surrogate model \citep{VarmaField2019}, the LIGO/Virgo collaboration derived the posterior distributions for the kick magnitude of GW190521 \citep{ligo2020new1} and found that the data are not very informative about the kick velocity of the merger remnant.

Using the results of our simulations, we estimate the average recoil kick imparted to the merger remnants in GWTC-1 and GWTC-2. We illustrate this in Figure~\ref{fig:all}, where we plot the probability distribution function of the average recoil kicks for the LIGO/Virgo population as a function of different $\chi_{\rm max}$. We find that the distributions are peaked at about $150\kms$, $250\kms$, $350\kms$, $600\kms$, for $\chi_{\rm max}=0.1$, $0.3$, $0.5$, $0.8$, respectively\footnote{Note that the maximum of $dP/d\log v_{\rm kick}$ might not coincide with the maximum of $dP/dv_{\rm kick}$, but for distributions with narrow peaks the difference is typically small.}.

Figure~\ref{fig:mqvkick} shows the average kick velocity imparted to the merger remnant of LIGO/Virgo BBH mergers as a function of their total mass and mass ratio. Different panels show different assumptions concerning $\chi_{\rm max}$. As expected, the larger the spin of the progenitors, the larger the recoil kick. GW190412 \citep{ligo2020}, a BBH merger with a mass ratio of nearly four-to-one, and GW190814 \citep{ligo2020b}, a merger between a BH and a compact object of about $2.5\msun$, have the lowest average recoil kicks owing to their small mass ratio. We represent in red circles the events whose $\chi_{\rm eff}$ and $\chi_{\rm fin}$ cannot reproduce the LIGO/Virgo estimated range for a given $\chi_{\rm max}$. We find that the estimated values of $\chi_{\rm eff}$ and $\chi_{\rm fin}$ of GW170729, GW190412, GW190519\_153544, and GW190620\_030421 can be reproduced if the spin of their progenitors was $\gtrsim 0.3$, and of GW190517\_055101 if the spin of their progenitors was $\gtrsim 0.8$\footnote{We have assessed the robustness of the results by running up to $\sim 10^7$ simulations for these GW events.}. This would imply that BHs could have not been born with low spins \citep{fullerma2019}. Alternatively, some of these events could be explained if at least one of the progenitors is already a second-generation BH, in agreement with the recent findings of \citet[][see also \citet{veske2020}]{kimball2020}.

We also estimate the probability that a given merger remnant would be retained within a star cluster of escape speed $v_{\rm esc}$, calculating the number of realizations for which $v_{\rm kick}<v_{\rm esc}$. The escape speed from the core of a star cluster is essentially determined by its mass and density profile: the more massive and denser the cluster is, the higher is the escape speed \citep[e.g.,][]{antorasio2016}. Open clusters, globular clusters, and nuclear star clusters have typical escape speeds of $\sim 1\kms$, $\sim 10\kms$, and $\sim 100\kms$ respectively\footnote{The escape speed may change over time depending on the details of its formation history and dynamical evolution \citep{rodr2020}.}.

We show in figure~\ref{fig:mqprobret} the probability to retain the merger remnant of LIGO/Virgo BBH mergers in environments with escape speed $v_{\rm esc}$. We plot results in different panels for different assumptions on $\chi_{\rm max}$. As in the previous figure, red circles represent events whose $\chi_{\rm eff}$ and $\chi_{\rm fin}$ cannot reproduce the LIGO/Virgo estimated range for a given $\chi_{\rm max}$. We find that only environments with escape speed $\gtrsim 100\kms$, as found in galactic nuclear star clusters as well as the most massive globular clusters and super star clusters, could efficiently retain the merger remnants of the LIGO/Virgo BBH population even for low progenitor spins ($\chi_{\rm max}=0.1$). In the case of high progenitor spins ($\chi_{\rm max}\gtrsim 0.5$), we conclude that only the most massive nuclear star clusters can retain the merger products.

\section{Conclusions}
\label{sect:concl}

The first and second Gravitational Wave Transient Catalogs have reported a total of $50$ confirmed events, providing a unique insight into their origin.

Assuming that LIGO/Virgo BBHs mergers were catalyzed by dynamical assembly and interactions in a dense star cluster, we compute the distribution of recoil kicks imparted to their merger remnants and estimate their retention probability within various astrophysical environments. We have found that the distributions of average recoil kicks are peaked at about $150\kms$, $250\kms$, $350\kms$, $600\kms$, for maximum progenitor spins of $0.1$, $0.3$, $0.5$, $0.8$, respectively. Only environments with escape speed $\gtrsim 100\kms$, as found in galactic nuclear star clusters as well as the most massive globular clusters and super star clusters, could efficiently retain the merger remnants of the LIGO/Virgo BBH population even for low progenitor spins ($\chi_{\rm max}=0.1$). In the case of high progenitor spins ($\chi_{\rm max}\gtrsim 0.5$), only the most massive nuclear star clusters can retain the merger products. 
Finally, we have found that some of the LIGO/Virgo events can be reproduced only if the BH progenitors were rapidly spinning. Alternatively, some of these events could be explained if at least one of the progenitors is already a second-generation BH.

As the sensitivity of current instruments improve and new detectors come online, interesting measurements of the recoil kicks imparted to merger remnants will soon become possible using robust estimates of the recoil kicks based on the analysis of the gravitational waveform \citep{GerosaMoore2016,ligo2020new1,VarmaIsi2020}. Nevertheless, our simple method still gives some interesting results for a few events (GW170729, GW190412, GW190519\_153544, GW190620\_030421, GW190517\_055101), which are in agreement with the results of recent works that use a comprehensive Bayesian framework \citep{kimball2020}.

\section*{Acknowledgements}

GF acknowledges support from CIERA at Northwestern University. This work was supported in part by Harvard's Black Hole Initiative, which is funded by grants from JFT and GBMF.\\
\ \\
\textit{Data Availability}\\
\ \\
The data underlying this article will be shared on reasonable request to the corresponding author.\\

\bibliographystyle{mn2e}
\bibliography{refs}

\end{document}